\documentstyle[11pt]{article}

\small
\begin{document}

\begin{titlepage}
\title{
{\bf The isospin independent spin-orbit force in the extended
Skyrme model } }
\author{ {\bf Abdellatif Abada}
\thanks  {email : "ABADA@nanvs2.in2p3.fr"}
\\
{\normalsize
Laboratoire de Physique Nucl\'eaire, CNRS/IN2P3 Universit\'e de Nantes, }\\
{\normalsize    2, Rue de la Houssini\`ere, 44072 Nantes Cedex 03, France }
}
\date{}
\maketitle
\begin{abstract}
By using the product ansatz as an approximation for the two-baryon system
we investigate the isoscalar nucleon-nucleon spin-orbit potential in an
extended Skyrme model including both fourth- and sixth-order terms.
As it is the case for the
Skyrme model, we still obtain the wrong sign for this interaction.
Nevertheless,
concerning the order of magnitude, the extended Skyrme model
provides a better agreement with phenomenological potentials,
as compared to the standard one.

\end{abstract}


\vskip 1cm
LPN 94-01 \hfill{February 1994}

\vskip 1cm
{\it Submitted to Phys. Lett. B }
\end{titlepage}

\newpage
The Skyrme model \cite{Sk}, recognized as the simplest chiral realization
of QCD at low energy and large $N_c$ \cite{Wi}, has been extensively used
to describe static and dynamical properties of baryons.
However, it has been shown recently \cite{Mo,AM}, through the
study of some properties of the nucleon, that one should not restrict
oneself
to the standard Skyrme model for the description of low-energy
hadron physics, but consider extensions of this model including higher order
terms in powers of the derivatives of the pion field.
In this letter we wish to study the effects of such additional terms
on the dynamical part of the nucleon-nucleon interaction.
We examine the influence of
the sixth-order term generated by $\omega$-meson exchange \cite{Jac}
on the nucleon-nucleon spin-orbit force, where
we focus on the isospin independent part of the interaction.
Several attempts to extract the isoscalar $N$-$N$ spin-orbit interaction from
the standard Skyrme model have given the wrong sign \cite{NR,ASW}.
Riska and Schwesinger \cite{RS} claimed that the inclusion of the
sixth-order term leads to the right sign.
However, these authors used a value for the corresponding parameter
which is in conflict with the experimental $\omega \to \pi \gamma$ width.

\vskip .5cm
\hskip .5cm {\Large {\bf 1.}}~
The effective Lagrangian density corresponding to the extended Skyrme model
we use can be expressed in terms of an SU(2) matrix field $U$
which characterizes the pion field. It reads
\begin{equation} \begin{array} {ll} \label{La}
{\cal L} = &{\displaystyle \frac {f_{\pi}^2 }{4}
{}~\hbox {Tr} ~(\partial_{\mu} U \partial^{\mu} U^{+})
+ \frac {1}{32 e^2}
{}~\hbox {Tr} ~\{~\left[ (\partial_{\mu} U) U^{+} , (\partial_{\nu} U) U^{+}
\right]^2  ~\} } \\
&{\displaystyle -\frac {1}{2} \frac{\beta_{\omega}^2}{m_{\omega}^2}
B_{\mu} B^{\mu}  + \frac {f_{\pi}^2 }{4} m_{\pi}^2
{}~\hbox {Tr} ~(U + U^{+} -2) }~
\end{array} \end{equation}
where $B_{\mu}$ is the baryon current \cite{Sk}
\begin{equation} \label{Bmu}
B^{\mu} = {\displaystyle \frac{1}{24 \pi^2}
\epsilon ^{\mu \nu \alpha \beta} \hbox {Tr} \left \{~
(\partial_{\nu} U) U^+ (\partial_{\alpha} U) U^+ (\partial_{\beta} U) U^+
{}~\right \} } .\end{equation}
The first term in Eq. (\ref{La}) corresponds to the
nonlinear $\sigma$-model, $f_{\pi}$ being the pion decay constant.
The second term, which is of fourth-order in
powers of derivatives of the pion field and parametrized by the
dimensionless factor $e$, was introduced by Skyrme in order to
stabilize the soliton. The third term is of order six in the derivatives
and can be derived from a local approximation of an effective model
with $\omega$ mesons \cite{Jac}. The constant $m_{\omega}$ is the
$\omega$-meson mass (782 MeV) while $\beta_{\omega}$ is a dimensionless
parameter related to the $\omega \to \pi \gamma $ width.
The last term in Eq. (\ref{La})
which is proportional to the square of the pion mass $m_{\pi}$  (139 MeV)
implements a small explicit breaking of chiral symmetry.

For the system of two interacting solitons we use the product
ansatz as suggested by Skyrme \cite{Sk2}. We also introduce rotational
dynamics to obtain the appropriate spin and isospin structure \cite{Ad1}.
Thus, the field configuration of the two-nucleon system separated by
a vector ${\bf r}$ reads:
\begin{equation} \label{pro}
U_{B=2} \equiv U_2({\bf x},{\bf r},A,B) =
A U_H({\bf x}-{\bf r}/2) C U_H({\bf x}+{\bf r}/2) B^+ ~,
\end{equation}
where $C = A^+ B$ and $A$ and $B$ are SU(2) matrices. To carry out a
simultaneous quantization of the relative motion of the two nucleons
and the rotational motion
we need to treat ${\bf r}, A$ and $B$ as collective coordinates. Hence
we make all these parameters (${\bf r}, A, B$) time dependent.
In Eq. (\ref{pro}),
$U_H$ is the commonly used SU(2) matrix for a single soliton with the
hedgehog ansatz:
\begin{equation} \label{hed}
U_H({\bf x}) = {\displaystyle \exp[~i \vec {\tau}.\hat {\bf x}
F(\vert {\bf x} \vert) ~] }~,
\end{equation}
where $F(\vert {\bf x} \vert)$ obeys the usual boundary conditions for
winding number one, and the $\tau_{a}$'s are the Pauli matrices.
The notation $\hat {\bf x}$ means ${\bf x}/ \vert {\bf x} \vert$~.
We have to note that the product configuration (\ref{pro}) is an approximation
which has inconveniences and advantages \cite{pas}.
We choose it for its relative
simplicity as compared to other two-baryon field configurations which can be
found in the literature. The region of validity of the product ansatz
corresponds to a separation $r$ much larger than 1 fm.

\vskip .5cm
\hskip .5cm {\Large {\bf 2.}}~
The spin-orbit potential will emerge due to a coupling between the relative
motion and the spins of the two nucleons. So we have to calculate the
kinetic energy
\begin{equation} \label{Ki}
K = {\displaystyle \int \hbox {d}^3 x ~{\cal K}(U_2) } ~,
\end{equation}
where ${\cal K}(U_2)$ is that part of the Lagrangian (\ref{La})
involving only quadratic time derivatives, in which $U$ is given by
Eq. (\ref{pro}).
The full expression of Eq. (\ref{Ki}) with respect to (${\bf r}, A, B$)
and their derivatives is very complicated and contains several terms which
do not need to be written in this letter.
They will be reported on a forthcoming
publication \cite{Aba2}. The isospin independent part $K_0$ of the
kinetic energy (\ref{Ki}) reads
\begin{equation} \label{K0}
K_0 = \frac{1}{2} \dot {\bf Q} {\bf M_0} \dot {\bf Q} ~,
\end{equation}
where
\begin{equation} \label{Q}
{\displaystyle
\dot {\bf Q}  = ( \dot {\bf r}, {\bf w}_+, {\bf w}_- ) }~.
\end{equation}
In Eq. (\ref{Q}), ${\bf w}_{\pm}$ are the sum and the difference of the
rotational velocities respectively:
$$
{\bf w}_{\pm} = -\frac{i}{2} \left(
\hbox{Tr} (\vec \tau A^+\dot A) \pm \hbox{Tr} (\vec \tau B^+\dot B) \right)~,
$$
and ${\bf M_0}$ is the $9\times 9$ mass matrix
\begin{equation} \label{mass}
 {\bf M_0}  =
\pmatrix{ \alpha (r)~I & -\beta(r)~ \epsilon \!\!/  & 0
\cr \beta(r)~ \epsilon \!\!/  & \gamma (r)~I & 0
\cr 0 & 0 & \gamma (r) I \cr}
\end{equation}
where $I$ is the 3$\times$3 identity matrix and
$ \epsilon \!\!/_{ij} \equiv \epsilon_{ijk} \hat {\bf r}_k $. The functions
$\alpha, \beta$ and $\gamma$ depend only on the relative distance
$r = \vert {\bf r} \vert$
and are determined from the chiral function $F$ (see Eq. (\ref{hed})).
We have to mention that the mass matrix ${\bf M_0}$ contains other terms,
such as a coupling between rotational velocities or a radial motion term,
which contribute in principle to $K_0$. However, we find these terms very
small and thus neglect them \cite{ASW,Aba2}. In this sense, the matrix
${\bf M_0}$ given by Eq. (\ref{mass}) should be considered as the dominant
contribution to the isospin independent kinetic energy (\ref{K0}).

Before identifying and extracting the spin-orbit potential from Eq. (\ref{K0})
one has to treat carefully the conversion from velocities to
canonical momenta. Indeed, one has to invert properly the mass matrix
${\bf M_0}$ in order to move from a Lagrangian formalism to a Hamiltonian one.
In terms of the canonical momenta
$${\displaystyle
{\bf P} = {\bf M_0} \dot{\bf Q} = ({\bf p_r},{\bf s},{\bf s}_-) ~,
}$$
where ${\bf p_r}$ is the conjugate momentum to ${\bf r}$, and
${\bf s} = {\displaystyle \frac{\vec{\sigma}_1}{2}
+ \frac{\vec{\sigma}_2}{2} } ,~
{\bf s}_- = {\displaystyle \frac{\vec{\sigma}_1}{2}
- \frac{\vec{\sigma}_2}{2} }  $  the conjugate momenta to
${\bf w}_+,~ {\bf w}_- $ respectively \cite{Ad1,Nym}
($\vec{\sigma}_i/2$ being the spin of the nucleon $i$, $i=1,2$)
, the kinetic energy
(\ref{K0}) becomes:
\begin{equation} \label{H0}
K_0 = \frac{1}{2} {\bf P} {\bf M_0}^{-1} {\bf P} ~.
\end{equation}
It is straightforward to invert the mass matrix (\ref{mass}). It's
expression reads:
\begin{equation}
{\bf M_0}^{-1} = {\displaystyle
\frac{1}{\alpha \gamma  -\beta^2} } \left[~~
\pmatrix{ \gamma~I & \beta~\epsilon \!\!/  & 0
\cr -\beta~\epsilon \!\!/  & \alpha ~I & 0
\cr 0 & 0 & (\alpha \gamma -\beta^2)/\gamma ~I \cr}
- {\displaystyle \frac {\beta^2}{\alpha \gamma} }~
\pmatrix{ \gamma ~\hat {\bf r}\!\!\!/ & 0 & 0
\cr 0 & \alpha ~\hat {\bf r}\!\!\!/& 0
\cr 0 & 0 & 0\cr} ~~\right ]
\end{equation}
where $\epsilon \!\!/$ has been defined above and
$\hat {\bf r}\!\!\!/_{ij} \equiv \hat {\bf r}_i\hat {\bf r}_j$.
By developing Eq. (\ref{H0}) we obtain
\begin{equation} \label{H02}
2 K_0 = {\displaystyle \frac{1}{\alpha \gamma - \beta^2} \left(~
\gamma ~{\bf p_r}^2 + \alpha ~{\bf s}^2  +
\frac{2\beta}{r} ~{\bf l}.{\bf s}
-\frac{\beta^2}{\alpha} ~({\bf p_r . \hat r})^2
-\frac{\beta^2}{\gamma} ~({\bf s. \hat r})^2  ~\right)
+ \frac{1}{\gamma} ~{\bf s_- . s_-} }
\end{equation}
where ${\bf l} = {\bf r} \times {\bf p_r}$ is the angular momentum.
Now we can easily extract  from Eq. (\ref{H02}) the isoscalar
spin-orbit force with the result:
\begin{equation} \label{pot}
V_{ls}(r) = {\displaystyle +\frac{\beta}{r~(\alpha \gamma - \beta^2)} }~.
\end{equation}

\vskip .5cm
\hskip .5cm {\Large {\bf 3.}}~
In order to compute the spin-orbit potential (\ref{pot}) we have to take a
set of parameters of the model, namely, $f_{\pi}, e$ and $\beta_{\omega}$.
The physically more sensible manner to choose the parameters is to remember
that the Skyrme model involves only meson fields, and where baryons emerge as
topological solitons. Therefore, since it is a meson theory, the parameters
of the model have to be fixed by fitting to the low-energy {\it meson}
observables and not to baryon data as it has been done in Ref. \cite{ASW}.
The experiment yields $f_{\pi} = 93$ MeV for the pion decay constant.
The dimensionless parameter $e$ can be determined from chiral
perturbation theory \cite{Gas}. The best set of
chiral low-energy constants has been determined by
Riggenbach {\it et al} \cite{Ri}, in analyzing the $K_{l4}$ decays.
As a consequence, one finds $e = 7.1 \pm 1.2$ \cite{Mo}.
Finally, the last parameter $\beta_{\omega}$ is obtained by fitting to the
$\omega \to \pi \gamma$ width, yielding $\beta_{\omega} = 9.3$ \cite{La}.
We have to mention that this set of parameters yields values of
the nucleon mass (including the Casimir effect), the
$\Delta$-$N$ mass splitting, the axial-vector coupling constant,
the Roper resonance, etc., close to the experimental situation \cite{Mo,AM}.

Our first result is that the isoscalar spin-orbit force obtained with these
parameters is {\it repulsive} and thus has the wrong sign as compared to the
experimental situation. This disappointing prediction of Skyrme models
is still not explained until now.
In order to compare our results to the data we consider therefore $-V_{ls}$
instead of $V_{ls}$ (c.f., Eq. (\ref{pot})).
Of course one can force the interaction to be
attractive by increasing the value of the parameter $\beta_{\omega}$ and
thus the sixth-order term dominates the second- and fourth-order terms.
This was done by the authors of Ref. \cite{RS} who used a value of
$\beta_{\omega} = 22.4$. However, this way to proceed is not
acceptable since this value of $\beta_{\omega}$ disagrees stongly
with the experimental $\omega \to \pi \gamma$ width.

We plot in Fig. 1 the behaviour of $-V_{ls}$ (\ref{pot}) with respect to the
nucleon-nucleon separation $r$ for
$[f_{\pi} = 93 $ MeV, $e = 7.1, \beta_{\omega} = 9.3]$ and compare
it to the standard Skyrme model prediction ($\beta_{\omega} = 0$). We also
plot in the same figure the corresponding part of the phenomenological
Bonn potential \cite{Bonn}.
{}From Fig. 1 we obviously see that the inclusion of the sixth-order term
in the Lagrangian (\ref{La}) improves considerably the prediction of the
isoscalar spin-orbit force
in the region of validity of the model ,i.e.,
$r \ge 1$ fm. It is expected that the agreement can not be perfect
due to several reasons.
One of them is the use of the approximation of the product ansatz
(\ref{pro}) to describe two-interacting nucleons.
A second reason is that the model used here (\ref{La})
should be regarded as a minimal extension of the Skyrme model.
Concerning the region of smaller $r$ (not shown on the figure) the
prediction of the model
differs completely from the phenomenological potential because this region
corresponds to processes of large momentum transfers where one should not
use effective Lagrangians but perturbative QCD.

\vskip .5cm
\hskip .5cm {\Large {\bf 4.}}~
In this letter we have derived the isospin independent spin-orbit interaction
from the extended Skyrme model including, in addition to the nonlinear $\sigma$
model, fourth- and sixth-order terms in powers of the derivatives of the pion
field.
Concerning the order of magnitude, we find a considerable improvement as
compared to the Skyrme model prediction
and thus, confirm the conclusions of Refs. \cite{Mo,AM}.
Indeed, the standard Skyrme model is insufficient for an accurate description
of baryon phenomenology and one should
generalize this model by including higher order terms in the chiral Lagrangian.
In spite of that, our major result is that, even with the sixth-order term
present, we still get the {\it wrong} sign for this force.
One should try to understand why the standard Skyrme model as well as the
extended one (\ref{La}) give the wrong sign and not mask the problem by
considering irrealistic values of the parameters \cite{RS}.
This difficulty with the isoscalar spin-orbit force seems to be similar to
that of missing central attraction. The problem concerning the lack of
attraction in the central potential has been solved by
considering effective Lagrangians which incorporate low mass mesons with
finite mass (the scalar meson having an important role) \cite{KV}.
Maybe one has to investigate in this way in order to correct the anomaly of
that sign \cite{Vinh}.

\vskip .5cm
{\bf Acknowledgments}

We are very grateful to D. Kalafatis, B. Loiseau, B. Moussallam, D. O. Riska,
R. Vinh Mau and N. R. Walet for helpful discussions and criticisms
and express our appreciation to D. Kalafatis and N. R. Walet
for a critical reading of the manuscript.

\vskip .5cm

\vskip .5cm
{\Large {\bf Figure captions } }
\vskip .5cm

{\bf FIG. 1.} The isoscalar spin-orbit potential (\ref{pot}) with a minus sign
in the case where the parameters fit to low energy {\it meson} data.
The dotted line corresponds to the Skyrme model (i.e., $\beta{\omega} = 0$),
while the full one corresponds to the extended model with
$\beta_{\omega} = 9.3$.
The dashed line is the corresponding term in the Bonn potential \cite{Bonn}.

\end{document}